\documentclass[11pt,a4paper]{article}

\usepackage[utf8]{inputenc}
\usepackage{amsmath}
\usepackage[mathcal]{euscript}
\usepackage{latexsym}
\usepackage{textcomp}
\usepackage{slashed}
\usepackage{tikz}
\usepackage{xspace}
\usepackage{setspace}
\usepackage[a4paper]{geometry}
\usepackage[numbers,sort&compress]{natbib}
\newcommand{\eprint}{}
\newcommand{\beq}{\begin{equation}}
\newcommand{\eeq}{\end{equation}}
\newcommand{\bqa}{\begin{eqnarray}}
\newcommand{\eqa}{\end{eqnarray}}
\newcommand{\bite}{\begin{itemize}}
\newcommand{\eite}{\end{itemize}}
\newcommand{\bd}{\begin{displaymath}}
\newcommand{\ed}{\end{displaymath}}
\newcommand{\bcen}{\begin{center}}
\newcommand{\ecen}{\end{center}}

\def\d#1{D_{#1}}
\def\tld#1{\tilde {#1}}

\newcommand{\nn}{\nonumber}
\newcommand{\nl}{\nonumber \\}

\def\Gosam{{{\sc GoSam}}}
\def\GOSAM{{{\sc GoSam}}}
\def\gosam{{{\sc GoSam}}}
\def\gosamtwo{{{\sc GoSam 2.0}}}

\def\form{{{\sc form}}}

\def\openloops{{{\sc OpenLoops}}}
\def\madloop{{{\sc MadLoop}}}

\def\samurai{{{\sc samurai}}}
\def\SAMURAI{{{\sc samurai}}}
\def\Sherpa{{{\sc Sherpa}}}

\def\cuttools{{{\sc CutTools}}}
\def\C++{{{\sc c++}}}

\def\Golem{{{\sc Golem95C}}}

\def\Ninja{{{\sc Ninja}}}
\def\MadLoop{{{\sc MadLoop}}}
\def\Madspin{{{\sc MadSpin}}}
\def\Pythia{{{\sc Pythia}}}

\def\amcnlomglong{{{\sc MadGraph5\_aMC@NLO}}}
\def\amcnlo{{{\sc MG5\_aMC}}}

\def\ttH{$t\bar{t}H$}
\def\ttyy{$t\bar{t}\gamma\gamma$}

\def\N{\mathcal{N}}

\begin{document}

\begin{center}

{\Large\bf Integrand reduction beyond one-loop calculations} \\
\vskip 18mm

{\bf Giovanni Ossola}\\
\small{ \tt gossola@citytech.cuny.edu}
\\[1em]
{\small {\it  New York City College of Technology, 
  City University of New York, \\ 300 Jay Street, Brooklyn NY 11201, USA} \\

  \vspace{0.22cm}

{\it The Graduate School and University Center, 
  City University of New York,  \\ 
  365 Fifth Avenue, New York, NY 10016, USA}
}
\end{center}

\vspace{1.3cm}

\begin{abstract}
\noindent In this presentation, we review the general features of integrand-reduction techniques, with a particular focus on their generalization beyond one loop. We start with a brief discussion of the one-loop scenario, a case in which integrand-reduction algorithms are well established and played over the past decade an important role in the development of automated tools for the theoretical evaluation of physical observables. 
The generalization of integrand-reduction techniques to all loops has been the subject of several efforts in the recent past, thus providing a better understanding of the universal properties of scattering amplitudes. The ultimate goal of these studies is the development of efficient alternative computational techniques for the evaluation of Feynman integrals beyond one loop. 
\end{abstract}

\vspace{1.9cm}

\begin{center}
{\it to appear in the proceedings of } \\

\vspace{0.2cm}

Loops and Legs in Quantum Field Theory\\
                 24-29 April 2016, 
                 Leipzig, Germany
\end{center}

\newpage

\section{Introduction}

Integrand-reduction techniques evolved enormously over the past decade. Since the one-loop 4-dimensional integrand reduction, also known as OPP method, introduced a new way of approaching the problem of the reduction of one-loop Feynman integrals~\cite{Ossola:2006us}, integrand reduction grew into a more general and variegated framework. Such advancements are due to several different authors and groups. A number of excellent review papers have been written on the subject and we refer the interested reader to them for more details and a more inclusive picture of the field~\cite{reviews}. 
 
In this talk, I will review some of the important steps along this evolution process, trying to underline the features that render integrand reduction a promising approach to study multi-loop scattering amplitudes. Most examples are taken from collaborative work in which I have been directly involved. They represent only a small part of the rich literature on the subject~\footnote{Among the many presentations given at this conference, see for example~\cite{ll16}.}.  

The presentation is organized as follows. I will start by defining the {\it integrand-level} approach to the reduction of scattering amplitudes, as compared with their {\it integral-level} description; I will then describe of the generalization of the integrand-level approach to higher loops, which was conveniently described using the language of multivariate polynomial division in algebraic geometry, and allowed to prove to a powerful theorem that shows the feasibility of such construction; After a brief excursus about the evolution of the numerical algorithms for one-loop calculations based on integrand reduction, I will give a short description of the \gosam\ framework and comment on the ongoing efforts towards its application beyond one loop. 

\section{Integrand level vs Integral level}
In order to introduce the notation and better define what we call the {\it integrand-level} reduction, let's consider a two-loop Feynman integral with n denominators:
\beq {\cal I} =  \int  dq \int dk\  {\cal A}({q,k}) =  \int  dq \int dk\  {{\cal N}({q,k}) \over 
           \d{1} \d{2} \ldots \d{n} } \label{eq:def} \, , \eeq
where $q$ and $k$ are the integration momenta. For the sake of this discussion, we don't need to specify at this stage whether such momenta are purely four-dimensional or regularized in $d=4 -2 \epsilon$ dimensions. The choice will become relevant later on when we will describe specific reduction algorithms. 

The {\it integral-level} approach leads to a description of  ${\cal I}$  in terms of Master Integrals {${\cal I}_i$}. The integral in Eq.~(\ref{eq:def}), can be rewritten as 
\beq
 {\cal I} = \int  dq \int dk\  {{\cal N}({q,k}) \over 
           \d{1} \d{2} \ldots \d{n} }  = c_0\  {{\cal I}_0} + c_1\ {{\cal I}_1} + \ldots + C_k \ {{\cal I}_k}
\eeq
for example by using tensorial reduction, exploring physical properties such as Lorentz invariance, or projecting the numerator to define convenient form factors. The initial integral is rewritten as a linear combination of  Master Integrals (MI) which are in principle general and easier to compute than the original integral at hands. If we are able to compute all MIs, the knowledge of the sets of coefficients which are in front of them is sufficient to solve the problem.

As an alternative path, we can manipulate the integrand {${\cal A}({q,k})$} of  Eq.~(\ref{eq:def}) and cast it to a more convenient form before tackling the integration. Such approach leads to the {\it integrand-level} reduction. 
The advantage of this approach is that integrands are much simpler to handle than integrals, since they are all rational functions, namely ratios of polynomials written in terms of integration variables and physical momenta and masses of the various particles involved in the scattering process. Moreover, the structure of the poles of {${\cal A}({q,k})$}, which play such an important role in field theory, is explicit in the integrand, as the set of zeros of the denominators. The question is how to use this knowledge to manipulate  {${\cal N}(q, k)$}, and therefore {${\cal A}(q, k)$}, to obtain a simpler decomposition after we put it back under the integral sign.

Of course the two approaches are deeply interconnected. Taking a look at the one-loop scenario  provides insights on their relation. In fact, at one loop, a general {\it integral-level} decomposition is well-known~\cite{Passarino:1978jh}: 
\bqa \nn \int dq\,  \frac{{\N(q)}}{\d{1} \d{2} \ldots \d{n}} &=& \sum_{\{i\}} {d_i} \int \frac{dq}{\d{i_1} \d{i_2} \d{i_3} \d{i_4}}
+\sum_{\{i\}} {c_i} \int \frac{dq}{\d{i_1} \d{i_2} \d{i_3}} \\
&+& \sum_{\{i\}} {b_i} \int \frac{dq}{\d{i_1} \d{i_2}}
+ \sum_i {a_i} \int \frac{dq}{\d{i}}
+ {{\rm R}}\, . \label{eq:pv}
\eqa
According to the decomposition in Eq.~(\ref{eq:pv}), any $n$-point Feynman integral, independently from the number of legs, can be written as a linear combination of 4-point, 3-point, 2-point, and 1-point scalar integrals, which therefore represent a complete set of MI at one loop, plus an additional ``constant'' term ${{\rm R}}$ know as the rational part. Since all scalar integrals are known and available in public codes~\cite{scalars}, the main problem in the evaluation of Feynman integrals lies in the evaluation of all the coefficients which multiply each MI. 

Eq.~(\ref{eq:pv}) can be used as a map to find the corresponding  {\it integrand-level} counterpart. If, for simplicity of notation, we consider purely four-dimensional integration momenta, where the rational part is absent, the {\it integrand-level} decomposition will have the form~\cite{Ossola:2006us}
{
\bqa \nn
{\N(q)} &=&
\sum_{\{i\}} 
\Big[
         {d_i} +
     {\tld{d_i}(q)} 
\Big]
\prod_{j \notin {\{i\}} } \d{j} 
     +
\sum_{\{i\}} 
\Big[
          {c_i} +
     {\tld{c_i}(q)}
\Big]
\prod_{j \notin {\{i\}} } \d{j}  \nl
     &+&
\sum_{\{i\}}
\left[
          {b_i}  +
    	{\tld{b_i}(q)}
\right]
\prod_{j \notin {\{i\}} } \d{j} 
     +
\sum_{i}
\Big[
          {a_i} +
     {\tld{a_i}(q)}
\Big]
\prod_{j \ne i} \d{j} \, ,
\label{eq:opp} \eqa
where all coefficients $d_i$, $c_i$, $b_i$, and $a_i$ are the same as in  Eq.~(\ref{eq:pv}). In order for the integrand decomposition of Eq.~(\ref{eq:opp}) to lead to the same result as the integral formula of  Eq.~(\ref{eq:pv}), the additional functions {$\tld{d}(q)$}, {$\tld{c}(q)$}, {$\tld{b}(q)$}, {$\tld{a}(q)$} should vanish upon integration: in the language of integrand reduction they are called {\it spurious terms}.
The general decomposition of Eq.~(\ref{eq:opp}) can be obtained algebraically by direct construction~\cite{delAguila:2004nf, Ossola:2006us}. All we need to do is rewriting $q$ in ${\N(q)}$ in terms of reconstructed denominators. The residual $q$ dependence, namely the terms that are not proportional to (products of) denominators, should vanish upon integration.  

After the identity of Eq.~(\ref{eq:opp}) is established, and the exact form of all spurious term has been determined, no further algebraic manipulation is needed. The functional dependence on the integration momentum is universal and process-independent, the only work required in order to compute the scattering amplitude is the extraction of all the coefficients. In the original integrand-level approach, the coefficients in front of the one-loop MIs are determined by solving a system of algebraic equations that are obtained by the numerical evaluation of the unintegrated numerator functions at explicit values of the loop-variable.
Such systems of equations become particularly simple when all expressions are evaluated at the complex values of the integration momentum for which a given set of inverse propagators vanish, that define the so-called quadruple, triple, double, and single cuts. This provides a strong connection between the OPP method and in general the integrand reduction techniques and generalized unitarity methods, where the on-shell conditions are imposed at the integral level. More details about algorithms for the implementation of integrand reduction will be provided in Section~\ref{sec:algo}. 

The idea of applying the integrand reduction to Feynman integrals beyond one-loop, pioneered in~\cite{Mastrolia:2011pr, Badger:2012dp}, has been the target of several studies in the past five years, thus providing a new
promising direction in the study of multi-loop amplitudes~\cite{Zhang:2012ce,Mastrolia:2012an,Badger:2012dv, Mastrolia:2013kca,intred2, Mastrolia:2016dhn}. By generalizing the language of Eq.~(\ref{eq:opp}), the numerator function in Eq.~(\ref{eq:def}) can be rewritten as~\cite{Mastrolia:2011pr}:
\beq
{\cal N}({q,k}) =
\sum_{i_1 < \!< i_8}^{n}
          \Delta_{i_1, \ldots, i_8}({q,k})
\prod_{h \ne i_1, \ldots, i_8}^{n} \d{h} 
+ \ldots 
+\sum_{i_1 < \!< i_2}^{n}
          \Delta_{i_1, i_2}({q,k}) 
\prod_{h \ne i_1 , i_2}^{n} \d{h} \, ,
\nn
\eeq
and thus accordingly
\beq
{\cal A}(q,k) =
\sum_{i_1 < \!< i_8}^{n}
         { \Delta_{i_1, \dots, i_8}({q,k}) \over 
           \d{i_1} \d{i_2} \ldots \d{i_8} } 
+
\sum_{i_1 < \!< i_7}^{n}
         { \Delta_{i_1, \dots, i_7 }({q,k}) \over 
           \d{i_1} \d{i_2} \ldots \d{i_7} } 
+
\ldots 
+
\sum_{i_1 < \!< i_2}^{n}
         { \Delta_{i_1, i_2}({q,k})  \over
           \d{i_1} \d{i_2} }  \label{eq:2loop} \, .
\eeq  
Unfortunately, beyond one loop, we cannot rely on the guidance of universal integral-level formulae in order to construct the integrand-level identity. Nevertheless the first  question that should be answered is the form of the polynomial residues  $ \Delta_{i_1, \ldots, i_m}$ appearing in the multipole expansion of Eq.~(\ref{eq:2loop}).
Like the one-loop case, their parametric form should be process-independent and determined once for all from the corresponding multiple cut. Unlike the one-loop case however, the basis of master integrals beyond one loop is not straightforward. Moreover, the splitting between ``spurious'' and ``physical'' terms in the residues is more tricky due to the presence irreducible scalar products (ISP), namely scalar products involving integration momenta that cannot be reconstructed in terms of denominators~\cite{Mastrolia:2011pr}.

As major milestone in this process, the determination of the residues at the multiple cuts has been systematized as a problem of multivariate polynomial division in algebraic geometry~\cite{Zhang:2012ce, Mastrolia:2012an}, which turned out to be a very natural language to describe the integrand-level decomposition. The use of these techniques allowed to apply the integrand decomposition not only at one loop, as originally formulated, but at any order in perturbation theory. Moreover, this approach confirms that the shape of the residues is uniquely determined by the on-shell conditions, without any additional constraint. 
In~\cite{Zhang:2012ce}, Yang Zhang presented an algorithm for the integrand-level reduction of multi-loop amplitudes of renormalizable field theories, based on computational algebraic geometry, and provided a Mathematica package, called BasisDet,  which allows for the determination of the various residues in the multi-pole decomposition of Eq.~(\ref{eq:2loop}). 

In~\cite{Mastrolia:2012an}, we proposed a general algorithm that allows to decompose any multi-loop integrand by means of a  powerful recurrence relation. In general, if the on-shell conditions have no solutions, the integrand is {\it reducible}, namely it  can be written in terms of lower point functions. One example of this class of integrands are the six-point functions at one loop, which are fully reducible to lower point functions, as well known for a long time~\cite{Kallen:1964zz}. When the on-shell conditions admit solutions, the corresponding residue is obtained dividing the numerator function modulo the Gr\"obner basis of the corresponding cut.
The {\it remainder} of the division provides  the {\it residue}, while the {\it quotients} generate integrands with less denominators. As a first application, we successfully reproduced in a straightforward manner all the residues which appear in the one-loop case.


The feasibility of the integrand reduction approach beyond one loop is guaranteed by the {\it Maximum Cut Theorem}~\cite{Mastrolia:2012an}. After labeling as {\it Maximum-cuts} the largest sets of denominators which can be simultaneously set to zero for a given number of loop momenta, the {\it Maximum Cut Theorem} ensures that the corresponding residues can always be reconstructed by evaluating the numerator at the solutions of the cut, since they are parametrized by exactly $n_s$ coefficients, where $n_s$ is the number of solutions of the multiple cut-conditions.  This theorem extends to all orders the features of the one-loop quadruple-cut in dimension four~\cite{Britto:2004nc,Ossola:2006us}, in which the two coefficients needed to parametrize the residue can be extracted by means of the  two complex solutions of the quadruple cut.


The recurrence algorithm can be applied both numerically and analytically~\cite{Mastrolia:2013kca}. In the \emph{ fit-on-the-cuts} approach, the coefficients which appear in the residues can be determined by evaluating the numerator at the solutions of the multiple cuts, as many times as the number of the unknown coefficients. This approach has been employed at one loop in the original integrand reduction~\cite{Ossola:2006us}, and has been generalized to all loops using the language of multivariate polynomial division. In the \emph{divide-and-conquer} approach~\cite{Mastrolia:2013kca}, the decomposition can be obtained analytically by means of polynomial divisions without requiring  prior knowledge of the form of the residues or the solutions of the multiple cuts, and the reduction algorithm is applied directly to the expressions of the numerator functions.

Very recently, a new simplified variant of the integrand reduction algorithm for multi-loop scattering amplitudes have been presented~\cite{Mastrolia:2016dhn} by Mastrolia,  Peraro, and Primo. The new algorithm exploits the decomposition of the integration momenta, defined in $d$-dimensions, in parallel and orthogonal subspaces with respect to the space spanned by the physical external momenta appearing in the diagrams. Non-physical degrees of freedom are integrated out by means of orthogonality relations for Gegenbauer polynomials, thus eliminating spurious integrals and leading to much simpler expressions for the integrand-decomposition formulae. This new fascinating approach has been presented at this conference by Pierpaolo Mastrolia, we refer the interested reader to his talk for further details.

\section{Integrand-reduction algorithms for one-loop amplitudes} \label{sec:algo}

\paragraph{Integrand-level Reduction in Four Dimensions} 
The integrand-reduction algorithm was originally developed in four dimensions~\cite{Ossola:2006us, Ossola:2007bb}, and implemented in the the code {\cuttools}~\cite{Ossola:2007ax}. The appearance of divergences in the evaluation of Feynman integrals requires the use of a regularization technique: in dimensional regularization the integration momentum is upgraded to dimension $d = 4 - 2 \epsilon$. Such procedure is responsible for the appearance of the rational part.
Following the OPP approach, there are two contributions to the rational term, which have different origins: the first contribution, called  ${\cal R}_1$, appears from the mismatch between the $d$-dimensional denominators of the scalar integrals and the $4$-dimensional denominators and can be automatically computed by means of a fictitious shift in the value of the masses~\cite{Ossola:2006us,Ossola:2007ax}. A second piece, called ${\cal R}_2$, comes directly from the $d$-dimensionality of the numerator function, and can be  recovered as tree-level calculations by means of \emph{ad hoc} model-dependent Feynman rules~\cite{rats}.

\paragraph{$D$-dimensional Integrand Reduction}  Since the rational term cannot be computed by operating in four dimensions, a very significant improvement have been achieved by performing the integrand decomposition directly in dimension $d = 4 - 2 \epsilon$\ rather than four~\cite{Ellis:2007br, Mastrolia:2010nb}, which indeed allows for the combined determination of all contributions at once. This approach requires to update of the polynomial structures in the residues to include a dependence on the extra-dimensional parameter $\mu$. These ideas, together the parametrization of the residue of the quintuple-cut in terms of the extra-dimensional scale \cite{Melnikov:2010iu} and the sampling of the multiple-cut solutions via Discrete Fourier Transform~\cite{Mastrolia:2008jb},  were the basis to the development of a new algorithm, called {\samurai}~\cite{Mastrolia:2010nb}.


\paragraph{Integrand Reduction via Laurent Expansion}  
If the polynomial dependence of the numerator functions on the loop momentum is known, all
coefficients in the integrand decomposition can be extracted by performing a Laurent expansion with respect to one of the free
parameters in the solutions of the various cuts implemented via polynomial division~\cite{Mastrolia:2012bu}.
This idea has been implemented in the \C++ library {\Ninja}~\cite{Peraro:2014cba}. Its use within the {\Gosam} framework  provided a significant improvement in the computational performance~\cite{vanDeurzen:2013saa}, both in terms of speed and precision, with respect to the previous algorithms, and has been employed in several challenging NLO  calculation, i.e. the evaluation of QCD corrections to $p p \to t {\bar t} H j $~\cite{vanDeurzen:2013xla} or to the associated production of a Higgs boson and three jets at the LHC in gluon fusion in the large top-mass limit~\cite{Cullen:2013saa}.
Thanks to recent work by Hirschi and Peraro, it is now possible to interface \Ninja\ to any one-loop matrix element generator that can provide the components of  loop numerator tensor~\cite{Hirschi:2016mdz}. This allowed, as a first example, to interface the  \Ninja\ library to \madloop~\cite{Hirschi:2011pa}, within \amcnlomglong~\cite{Alwall:2014hca}. A very detailed numerical analysis showed that \Ninja\  performs better that other reduction algorithms both in speed and numerical stability~\cite{Hirschi:2016mdz}.

\section{\gosamtwo\ for one-loop calculations and beyond}

The {\gosam} framework~\cite{Cullen:2011ac} combines automated Feynman diagram generation and algebraic manipulation~\cite{algebra}, with tensorial decomposition and integrand reduction, to allow for the automated numerical evaluation of virtual one-loop contribution to any given process. After the generation of all Feynman integrals contributing to the selected process, the virtual corrections can be evaluated using the integrand reduction via Laurent expansion~\cite{Mastrolia:2012bu} provided by {\Ninja}, which is the default choice, or the $d$-dimensional integrand-level reduction method, as implemented in \SAMURAI~\cite{Mastrolia:2010nb}, or alternatively the tensorial decomposition provided by {\Golem}~\cite{Binoth:2008uq}. 
The only task required from the user is the preparation of an input file for the generation of the code and the selection of the various options, without having to worry about  the internal details. 

The computation of physical observables at NLO accuracy, such as cross sections and differential distributions, requires to interface \gosam{} with other tools that can take care of the computation of the real emission contributions and of the subtraction terms, needed to control the cancellation of IR singularities, as well as the integration over phase space.

As an example of application, a new interface that was recently developed between the multipurpose Monte Carlo tool \amcnlomglong\ and \gosam~\cite{vanDeurzen:2015cga}. 
In order to validate the interface several cross checks were performed.  The
loop amplitudes of \gosam{} and \madloop{} were compared for
single phase space points and also at the level of the total cross
section for a number of different processes, as presented in a
dedicated table in~\cite{hansthesis}. Furthermore, for $pp \rightarrow
t\bar{t}\gamma\gamma$, a fully independent check was also performed by
computing the same distributions using \gosam{} interfaced
to \Sherpa{} (see Figure~\ref{fig:ttyy}). 

\begin{figure}[ht] 
\bcen
{\includegraphics[width=7.5cm]{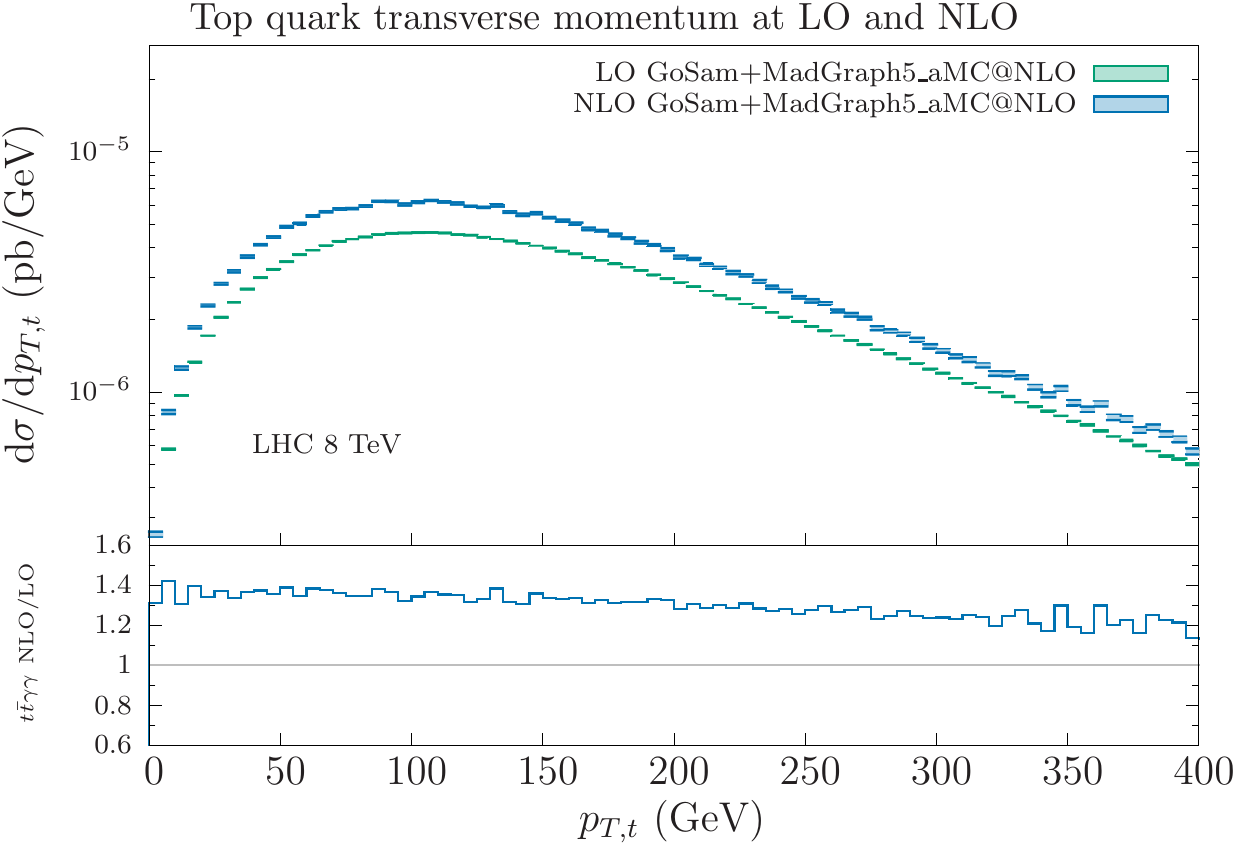}} 
\hspace{0.31cm}
 {\includegraphics[width=7.5cm]{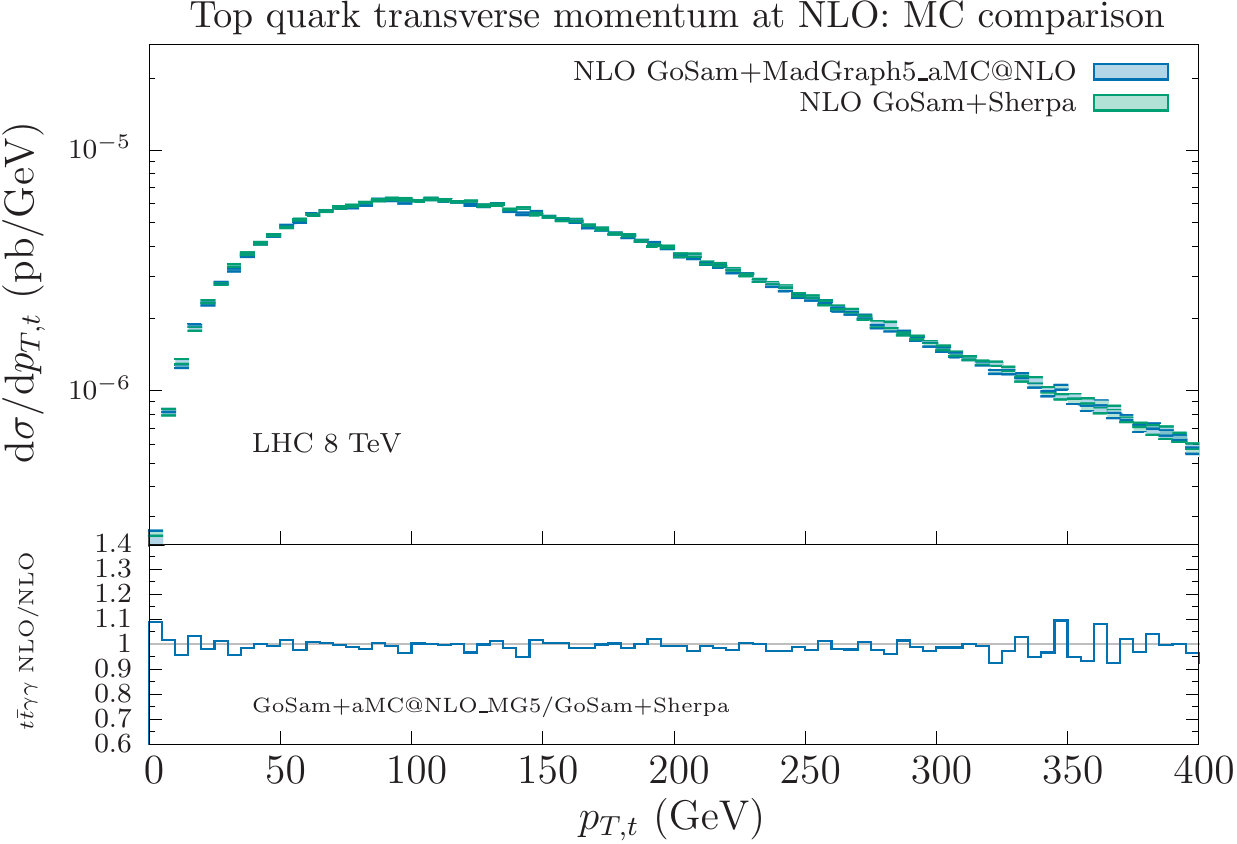}}\\
\ecen
\caption{Transverse momentum of the top quark in $ p p \to t \bar t \gamma \gamma$  for the LHC at 8 TeV: 
LO and NLO distributions for the transverse momentum of the top quark (left) and NLO comparison between \GOSAM{}+\amcnlo\ and \GOSAM{}+\Sherpa\ (right).  }
\label{fig:ttyy}
\end{figure}

As an application of this novel framework, we computed the NLO QCD corrections to $pp \to$ \ttH{} and $pp \to$ \ttyy{} matched to a parton shower~\cite{vanDeurzen:2015cga}. The study is performed using NLO predictions for \ttH\ and continuum \ttyy{} production. The top and anti-top quarks are subsequently decayed semi-leptonically  with \Madspin~\cite{Frixione:2007zp}, taking into
account spin correlation effects, and then showered and hadronised by means of  \Pythia~8.2~\cite{Sjostrand:2014zea}.
We compared several distributions to disentangle the two processes and focused in particular on
observables designed to study spin correlation effects. While NLO corrections are sizable and provide a clear reduction of theoretical uncertainties, they only mildly distort the shape of the various distributions.
\begin{figure}[h!] 
\bcen
{\includegraphics[width=6.5cm]{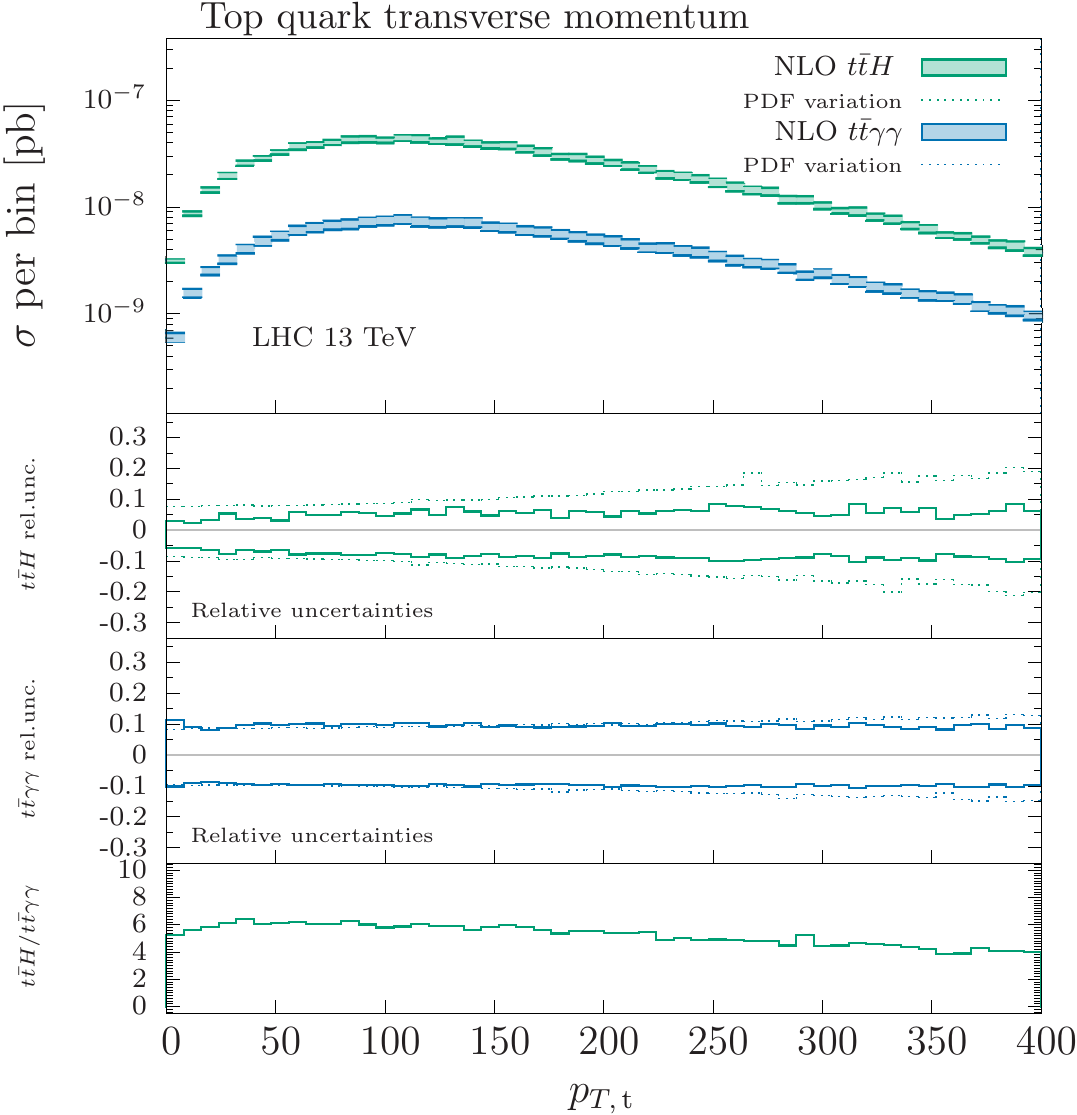}} 
\hspace{0.7cm}
 {\includegraphics[width=6.5cm]{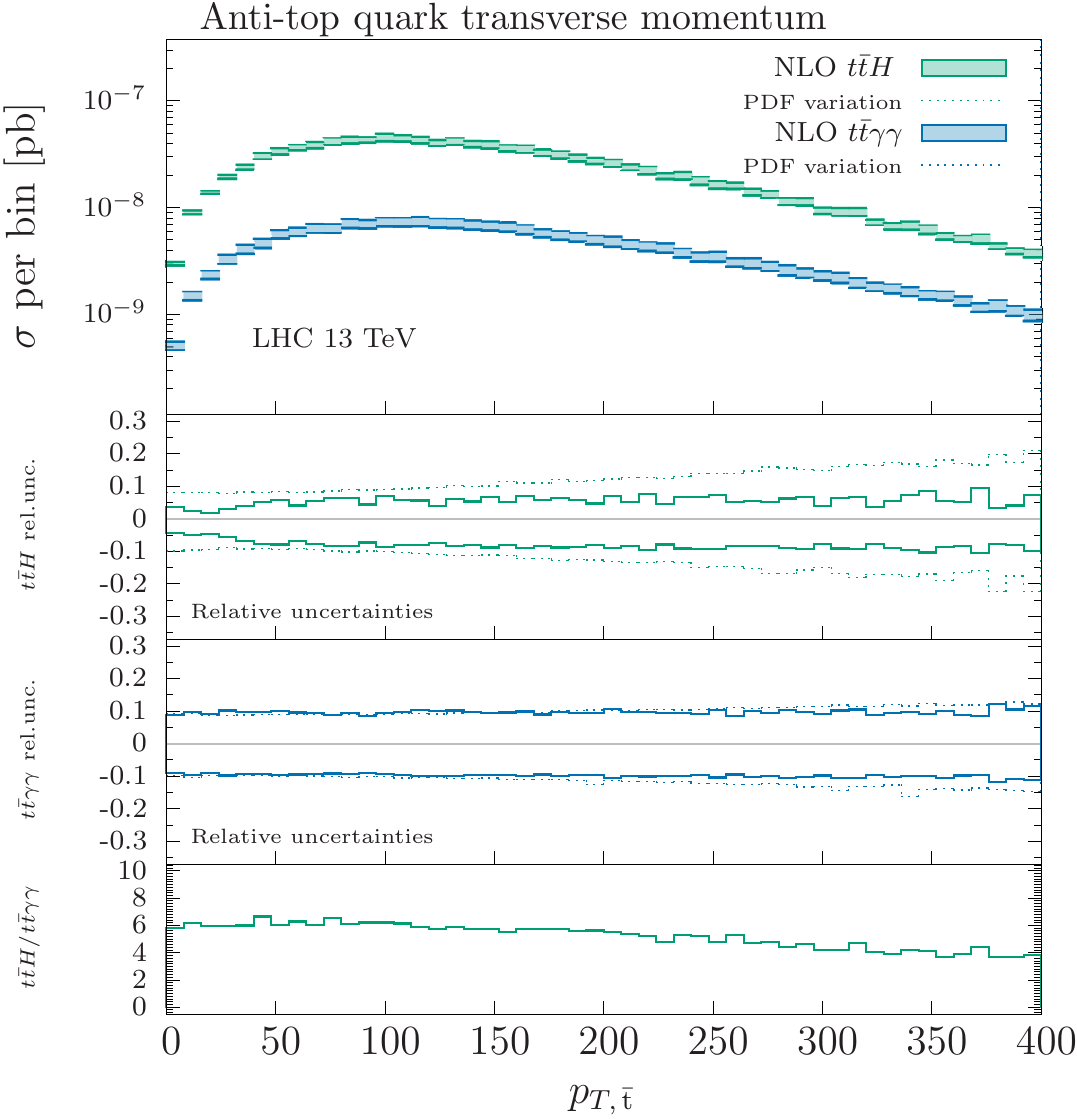}}\\
\ecen
\caption{Transverse momentum of the top and anti-top quark in $ p p \to t \bar t \gamma \gamma$  for the LHC at 13 TeV.}
\label{fig:ttyy2}
\end{figure}

While \gosam\ was initially designed and developed to compute one-loop virtual contributions needed by NLO predictions, several of it features can be adapted and extended to address specific tasks needed by higher order calculations.  Concerning the generation virtual two-loop matrix elements, the routines in \gosam\  have been extended~\cite{Borowka:2016agc} to produce the full list and expressions for all two-loop Feynman diagrams contributing to any process: as for the one-loop case, the code depicts all contributing diagrams as output on file, takes care of the algebra by means of \form, and projects the expressions over the appropriate tensor structures, to extract the form factors. Interfaces to codes for the reduction to master integrals are also available. The code had been successfully employed for the evaluation of the two-loop virtual amplitudes needed by the evaluation of Higgs boson pair production in gluon fusion at NLO, where relevant master integrals have been computed numerically by means of SecDec-3.0~\cite{Borowka:2015mxa}. More details about this important result have been presented at this conference by Stephen Jones and Matthias Kerner.

As a last application beyond NLO, \gosam\ has been used for the evaluation of  $pp\ \to t\bar{t}H$ and $pp\ \to t\bar{t}W$ at approximate NNLO in QCD~\cite{Broggio:2015lya}.
In these papers, approximate formulas were obtained by studying
soft-gluon corrections in the limit where the invariant mass of the
final state approaches the partonic center-of-mass
energy. No assumptions are made about  the invariant mass of the final state.
 The approximate NNLO corrections are extracted from the
 perturbative information contained in a soft-gluon resummation formula valid to NNLL accuracy, whose
 derivation is based on SCET (for a recent review, see~\cite{Becher:2014oda}). 
 
 The soft-gluon resummation formula for this process contains three essential ingredients, all of which are matrices in the
 color space needed to describe four-parton scattering: a hard
 function, related to virtual corrections; a soft function, related
 to real emission corrections in the soft limit; and a soft
 anomalous dimension, which governs the structure of the all-order
 soft-gluon corrections through the renormalization group (RG).
Of these three ingredients, both the NLO soft function~\cite{Ahrens:2010zv} and NNLO soft anomalous dimension~\cite{Ferroglia:2009ep} needed for NNLL resummation in processes involving two massless and two massive partons can be adapted directly to $t\bar{t}H$ and $t\bar{t}W$ production.
The NLO hard function is instead process dependent and it was evaluated by using a modified version of the one-loop providers \Gosam, \MadLoop, and \openloops~\cite{Cascioli:2011va}.

\section{Conclusions}

Integrand-reduction techniques played an important role in the automation of NLO calculations. Algorithms such as the four-dimensional integrand-level OPP reduction, $D$-dimensional integrand reduction, integrand reduction via Laurent expansion, embedded in multi-purpose codes interfaced within Monte Carlo tools, allowed to compute cross sections and distributions for a wide variety of processes at NLO accuracy, as needed by the LHC experimental collaborations.

However, integrand reduction did not merely provide a set of numerical and computational algorithms to extract coefficients, but a different approach to
scattering amplitudes, based on the study of the general structure of the integrand of Feynman integrals. The advances during past decade also showed that a better understanding of the mathematical properties of scattering amplitudes goes together with the ability of developing efficient algorithms for their evaluation. Moreover, there is still room for improvement, even at NLO.

Will integrand reduction be competitive at NNLO?  The challenges and additional complexity presented by NNLO calculations required an extension of integrand reduction techniques to go beyond the current understanding. The language of algebraic geometry provided an ideal framework to determine the functional form of all residues, and attempts of optimizing and reducing the number of terms generated by the integrand decomposition have been presented and will be soon implemented in computational codes. More work is still needed, but integrand reduction might provide an alternative path for NNLO processes with more than two particles in the final state.

\paragraph{Acknowledgments} The results presented in this paper are the outcome of the team work with several talented and motivated collaborators. I would like to thank the present and former members of the \gosam\ Collaboration for their many contributions. I am particularly indebted with Pierpaolo Mastrolia, Gionata Luisoni, and Tiziano Peraro, for everything I learned from our discussions and common projects. I would also like to thank Alessandro Broggio, William Bobadilla, Andrea Ferroglia, Valentin Hirschi, Amedeo Primo, and Ray Sameshima for stimulating discussions on a wide variety of topics over the past year. Work supported in part by the National Science Foundation under Grants  PHY-1068550 and PHY-1417354 and by the PSC-CUNY Awards No. 67536-00 45 and No. 68687-00 46.


\end{document}